\documentclass[aps,twocolumn,prl,showpacs,preprintnumbers,amsmath,amssymb]{revtex4}

\usepackage{graphicx}
\usepackage{dcolumn}
\usepackage{bm}
\usepackage{txfonts}
\usepackage{color}
\usepackage{amsmath}
\usepackage{amsfonts}
\usepackage{amssymb}

\begin{document}
\preprint{APS/123-QED}

\title{Anomalous effect of doping on the superconducting state of CeCoIn$_5$ in high magnetic fields}
\author{Y. Tokiwa$^{1,2}$}
\author{R. Movshovich$^1$}
\author{F. Ronning$^1$}
\author{E. D. Bauer$^1$}
\author{A. D. Bianchi$^3$}
\author{Z. Fisk$^4$}
\author{J. D. Thompson$^1$}
\affiliation{$^1$Los Alamos National Laboratory, Los Alamos, New Mexico 87545, USA}

\affiliation{$^2$I Physikalisches Institut, Georg-August Universit\"{a}t G\"{o}ttingen, G\"{o}ttingen 37077,
Germany}

\affiliation{$^3$D\'{e}partement de physique Montr\'{e}al, Universit\'{e} de Montr\'{e}al, QC, H3C 3J7
Canada }

\affiliation{$^4$Department of Physics, University of California at Irvine, Irvine, California 92697,
USA}

\date{\today}

\begin{abstract}
We investigated the effect of electron and hole doping on the high-field low-temperature superconducting
state in CeCoIn$_5$ by measuring specific heat of CeCo(In$_{\rm 1-x}$M$_{\rm x}$)$_5$ with M=Sn, Cd and
Hg and $x$ up to 0.33\%  at temperatures down to 0.1\,K and fields up to 14\,T. Although both Cd- and
Hg-doping (hole-doping) suppresses the zero-field $T_c$ monotonically, $H_{c2}$ increases with small
amounts of doping and has a maximum around $x$=0.2\% (M=Cd). On the other hand, with Sn-doping
(electron-doping) both zero-field $T_c$ and $H_{c2}$ decrease monotonically. The critical temperature
for the high-field low-temperature superconducting state correlates with $H_{c2}$ and $T_c$,
which we interpret in support of the superconducting origin of this state.

\end{abstract}

\maketitle

Magnetic field's coupling to spins of the electrons of a superconducting pair (Cooper pair) destroys
superconductivity via the Pauli limiting effect~\cite{clogston:prl-62}. In the normal state, magnetic
field splits the spin-up and spin-down states (Zeeman splitting), and the system can lower its free
energy by preferentially populating the lower energy level (spin-up band). This also gives rise to the
Pauli susceptibility. On the other hand, the energy of a spin-singlet SC state can not be affected by
Zeeman effect, because electrons pair up with opposite spins, i.e. there is an equal number of
superconducting spin-up and spin-down electrons. When the energy gain from Zeeman effect in the normal
state exceeds the SC condensation energy with increasing magnetic field, superconductivity is destroyed
at a Pauli-limiting field $H_{\rm P}$. The field's coupling to the orbital motion of the superconducting
electrons also can destroy superconductivity at a characteristic field $H_{\rm c2}^{\rm 0}$. For
strongly Pauli limited superconductors (when the Maki parameter $\alpha=\sqrt{2}H_{c2}^0/H_P$ is large)
a number of peculiar properties were anticipated theoretically in 1960's. The SC transition was
predicted to become first order at high fields, when $T_c$ is suppressed below $0.56
T_c$($H$=0)~\cite{maki:ptp-64,SARMAG:theiue}. The Zeeman effect was also expected to lead to a spatially
inhomogenious SC state, stable at high fields and low temperatures, proposed by Fulde and
Ferrell~\cite{fulde-ferrell:pr-64} and Larkin and Ovchinnikov~\cite{larkin-ovchinnikov:jetp-64} (FFLO).

The discovery of a strongly Pauli-limited superconductor CeCoIn$_5$ has triggered numerous
investigations of its peculiar SC properties~\cite{petrovic:jpcm-01,movshovich:prl-01}. CeCoIn$_5$ is SC
below $T_c$=2.25\,K with a large Maki parameter $\alpha$, which is  anisotropic with respect to the
direction of the magnetic field and ranges between 3.5 ($H \parallel$ [001]) and 4.5 ($H \parallel$
[100])~\cite{bianchi:prl-02}. CeCoIn$_5$ shows a first order SC transition in a bulk superconductor via
specific heat and magnetization anomalies at high fields~\cite{bianchi:prl-02,tayama:prb-02}, in accord
with the above mentioned theoretical expectations. When a high magnetic field close to $H_{c2}$ is
applied within the basal ({\it a-b}) plane of the tetragonal crystal structure of CeCoIn$_5$, specific
heat shows an additional anomaly within the superconducting state into a High Field Low Temperature
(HFLT) phase, originally proposed to be a realization of a long-searched-for FFLO
state~\cite{bianchi:prl-03a,radovan:nature-03}.

In this Rapid Communication, we show that very small amounts of both magnetic (Cd and Hg) and
non-magnetic (Sn) doping have dramatic effects on both gross superconducting properties ($H_{c2}$) and
the evolution and stability of the HFLT phase in CeCoIn$_5$. Our data, in particular the broadening of
the specific heat anomaly, suggest that the underlying origins of this phase are indeed of the FFLO
nature.

The HFLT transition itself has been confirmed by penetration depth~\cite{martin-prb05}, thermal
conductivity~\cite{capan-prb04}, ultrasound velocity~\cite{watanabe-prb04},
magnetostriction~\cite{correa-prl07}, and nuclear magnetic resonance
(NMR)~\cite{kakuyanagi-prl05,mitrovic-prl06,kumagai-prl06} measurements (for a recent review, see
Ref.~\cite{matsuda:jpsj-07}). Importantly, NMR investigation~\cite{YoungBL:Micefm} revealed a long range
antiferromagnetic (AFM)order within the HFLT state~\cite{YoungBL:Micefm}. Further, magnetic Bragg peaks
with an incommensurate propagation vector $\vec{Q} = (0.44,0.44,0.50)$, was found within the HFLT SC
phase by neutron scattering~\cite{Kenzelman-Science09}, and that phase was named by the authors $Q$-phase, in reference to the observed $\vec{Q}$. The presence of the long range AFM order revealed by the
NMR and neutron scattering experiments gave rise to an alternative possibility of magnetism being the
driving force of the phase transition into the HFLT state. Our present doping studies test this
hypothesis.

\begin{figure}[t]
\includegraphics[width=\linewidth,keepaspectratio]{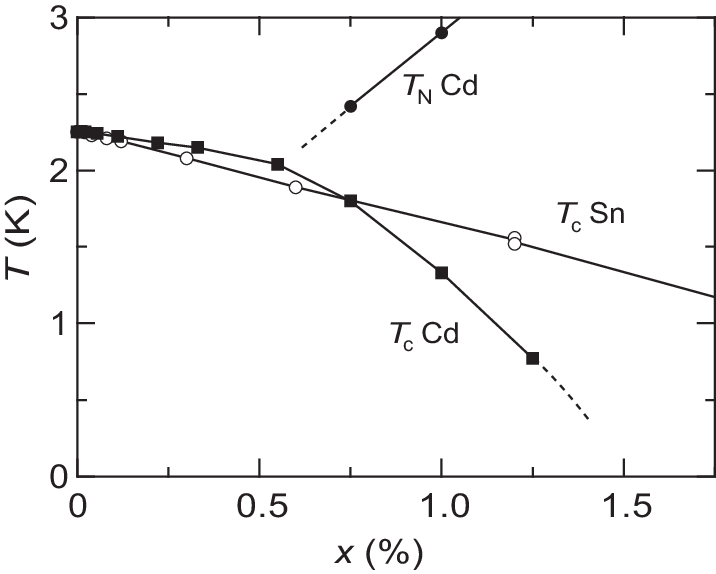}
\caption{Zero-field superconducting transition temperature, $T_c$, of CeCo(In$_{\rm 1-x}$Cd$_{\rm
x}$)$_5$ and CeCo(In$_{\rm 1-x}$Sn$_{\rm x}$)$_5$~\cite{bauer-prb06} as a function of concentration,
together with antiferromagnetic transition temperature, $T_{\rm N}$, of CeCo(In$_{\rm 1-x}$Cd$_{\rm
x}$)$_5$.} \label{}
\end{figure}

Figure~1 shows evolution of $T_c$ and $T_N$ for Cd-doped and $T_c$ for Sn-doped~\cite{bauer-prb06}
CeCoIn$_5$ at zero-field. Both Cd- and Sn-doping suppress the SC state monotonically and Cd-doping
induces AFM order above 0.5\,\%. Hg-doping has essentially the same effect as
Cd-doping.~\cite{bauer-physicab08}

Since Cd-doping induces AFM-order, one would naively expect it to also stabilize the HFLT phase with its
long range AFM order. However, the HFLT phase was found to be extremely fragile with respect to Cd and
Hg impurities~\cite{tokiwa-prl08}. Only 0.05\% of Hg on In-sites destroys the HFLT phase. Furthermore,
Fig.~1 shows that the AFM order speeds up the suppression of $T_c$ with a steeper slope above 0.5\,\%,
implying a competition between the two states, while the magnetically ordered HFLT phase appears to need
superconductivity for its existence.

The AFM order in CeCoIn$\rm_5$, induced by Cd doping above 0.5\,\%, potentially may be related to the
AFM order within the HFLT phase. To untangle the effect of suppression of the HFLT phase at low
impurities concentration from that of inducing long range magnetic order at higher levels of Hg and Cd
doping, it is important to perform complimentary investigations of the response of the HFLT phase to a
non-magnetism-inducing impurity. Doping with Sn achieves just that, as Sn-doping was shown to suppress
SC without driving the system into an AFM ordered state~\cite{bauer-prl05}.

Specific heat measurements were performed on single plate-like samples with a typical weight of 1-3 mg. Initial sample characterization via micro-probe analysis, using wavelength dispersive
spectroscopy, showed uniform distribution of the dopants. Specific heat was measured in a dilution
refrigerator with a superconducting magnet, employing the quasi-adiabatic heat pulse method.

We first address the effects of doping on the gross features of the phase diagram of CeCoIn$_5$. In zero
field (see Fig.~1) the low doping behavior for Cd and Sn doping is similar, reflecting impurity pair
breaking effects. At high field the response of CeCoIn$_5$ to small amounts of hole and electron doping
is strikingly different. Figure~2 shows $H$-$T$ phase diagrams of Cd-doped CeCoIn$_5$ for the field
along [001] and [100]. $H_{\rm c2}$ increases for both $H\parallel$ [001] and [100]. The zero-field
$T_c$ monotonically decreases with Cd doping, and increase of $H_{\rm c2}$ leads to the crossing of SC
phase boundaries for the undoped and Cd-doped CeCoIn$_5$ (Fig.~2b), indicating more stable SC state at
high fields in the Cd-doped compounds.  Extrapolated $H_{\rm c2}$ to zero temperature is shown in the
insets. $H_{\rm c2}$ exhibits a maximum at $x = 0.22$\,\% for $H\parallel$[100], while it increases
monotonically up to $x = 0.33$\,\% for $H\parallel$[001].

\begin{figure}[t]
\includegraphics[width=\linewidth,keepaspectratio]{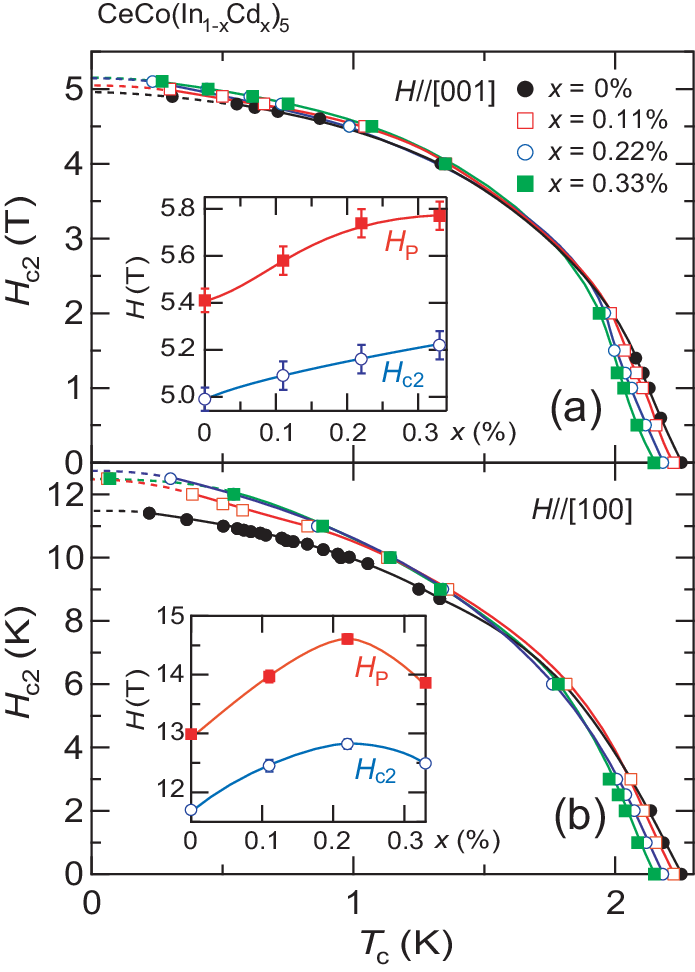}
\caption{(Color online) $H$-$T$ phase diagrams of CeCo(In$_{\rm 1-x}$Cd$_{\rm x}$)$_5$ with $x$=0, 0.11,
0.22 and 0.33\,\% for $H\parallel$[001] (a) and $H\parallel$[100] (b). Solid lines are guides to eye.
Insets: Experimental upper critical field $H_{\rm c2}$ and deduced Pauli-limiting field $H_{\rm P}$ as
a function of Cd-doping concentration.} \label{}
\end{figure}

The orbital limiting field, $H_{c2}^0 = 0.7 T_c dH_{c2}/dT|_{T=T_c}$, for $H\parallel$ [100] is 37, 37,
36 and 39$\pm$1\,T for x=0, 0.11, 0.22 and 0.33\,\%, respectively. It is rather independent of Cd
concentration, and can not be responsible for the increase in $H_{c2}(0)$. The Pauli-limiting field is
estimated from the orbital limiting field and experimental $H_{c2}$, using the results of the numerical
calculations in Ref~\cite{gruenberg:prl-66}. The resulting $H_{\rm P}$ is shown in the insets of Fig.~2,
and it follows the same trend as $H_{\rm c2}$. The Pauli-limiting field $H_{\rm
P}$=$\sqrt{2}\Delta/g\mu_{\rm B}$, where $\Delta$ is the SC energy gap at zero temperature and $g$ is
the gyromagnetic ratio. Assuming that $\Delta$ is proportional to zero-field $T_c$, the relative change
of $g$ as a function of doping is $g$($x$)/$g$(0)=$[T_c(x)/H_P(x)]/[Tc(0)/H_P(0)]$ (not shown). As
$H_{\rm P}$ exhibits a maximum at 0.22\,\% for $H\parallel$[100], $g$ has a minimum, with the reduction
of 14\,\% from $g(0)$. The $g$-factor as a function of pressure was found previously to increase from
0.632 at ambient pressure to 0.6365 at 0.45 \,GPa (a change of less then a percent), and then drop to
0.554 at 1.34\,GPa.~\cite{miclea-prl06,knebel-jpsj09}. In the earlier zero-field study~\cite{pham:prl-06}, it was concluded that Cd-doping acts as negative pressure, based on the finding that Cd-doping suppress SC and induces AFM order, while pressure suppresses AFM and can drive AFM ordered
CeCo(In$_{\rm 1-x}$Cd$_{\rm x}$)$_5$ back to non-magnetic SC ground state. Here we note that Cd doping of 0.22\,\% corresponds to an effective negative pressure of only 0.28\,GPa~\cite{pham:prl-06}. Therefore, the relative change of the $g$-factor with Cd doping is
at least an order of magnitude greater than can be accounted for by the pressure effect alone.

\begin{figure}[t]
\includegraphics[width=\linewidth,keepaspectratio]{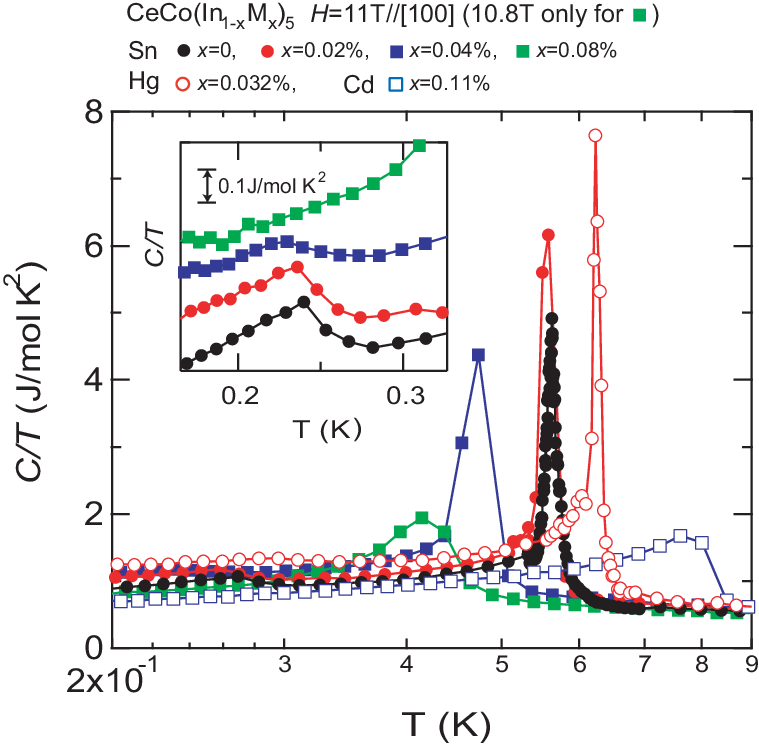}
\caption{(Color online) Specific heat divided by temperature, $C/T$, for CeCo(In$_{\rm 1-x}$T$_{\rm
x}$)$_5$ (T=Sn, Hg and Cd) as a function of temperature at 11\,T and at 10.8\,T for T=Sn and
$x$=0.08\,\%. Inset: Expanded plot of $C/T$ around HFLT phase transition for T=Sn. Data are shifted
vertically for clarity.} \label{}
\end{figure}

We now address the effect of small levels of electron and hole doping on the HFLT phase. We will show that Sn (electron) doping has both similar and opposite effects, to those of Cd and Hg (hole) doping, on
the superconducting state of CeCoIn$_5$. However, the data are consistent and can be understood within
one picture. 

Figure~3 shows specific heat divided by temperature, $C/T$, of CeCo(In$_{\rm 1-x}$T$_{\rm
x}$)$_5$ (T=Cd, Hg and Sn) at 11\,T, and for T=Sn at $x$=0.08\,\% at 10.8 T, for doping concentrations
up to 0.11\,\%.  The HFLT-phase anomaly, displayed in the inset of Fig.~3, broadens with increasing
concentration (as in the case of Cd and Hg doping) and disappears at $x$=0.08\,\%. The critical
Sn-concentration to destroy the specific heat anomaly associated with the HFLT phase, 0.08\,\% at most,
is similar to 0.05\,\% for Hg doping. These results show that the HFLT phase is extremely fragile with
respect to impurities, regardless of whether the same impurities at higher doping levels stabilize AFM
ground state or not. The inter-impurity distance at the critical doping concentrations of 0.05\% (the
characteristic length scale of the HFLT state) is $\sim$40\,$\AA$, which is comparable to the SC
coherence length (70\,$\AA$), supporting the SC origin of the HFLT state, instead of an AFM one.

The effects on the normal-SC phase boundary at high fields for the two types of dopants in the
low-doping regime are dramatically different. For Sn-doping, $T_c$($H$=11\,T) shifts monotonically to
lower temperature, contrary to the effect of Cd and Hg doping discussed above. Electron doping (with Sn)
suppresses $H_{c2}$ roughly by the same amount that Cd and Hg increase it for the same amount of doping
(see Fig. 5), supporting the picture of charge doping (Fermi surface effect) being the dominant driver
of the change of $H_{c2}$.

\begin{figure}[t]
\includegraphics[width=\linewidth,keepaspectratio]{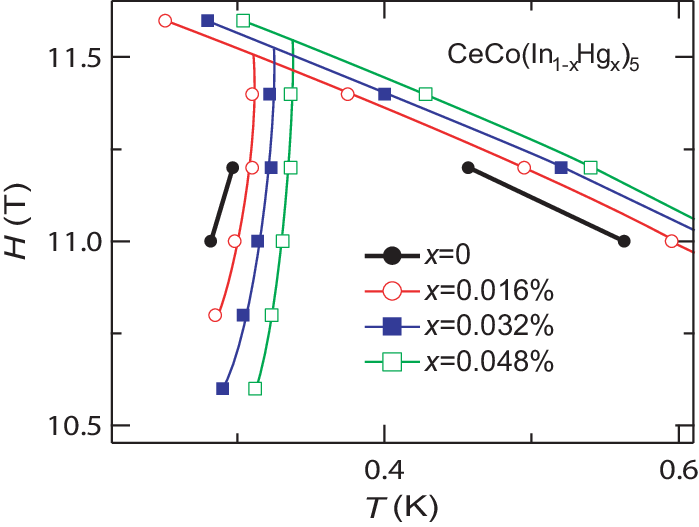}
\caption{(Color online) $H$-$T$ phase diagram of CeCo(In$_{\rm 1-x}$Hg$_{\rm x}$)$_5$ for the HFLT phase
region of SC state. Solid lines are guides to eye.} \label{}
\end{figure}

The response to charge doping of the superconducting critical field correlates well with that of the HFLT phase. A recent study has
shown that $T_{\rm HFLT}$ increases slightly with Hg-doping, while the specific heat anomaly associated
with the transition into the HFLT state broadens and is quickly destroyed~\cite{tokiwa-prl08}. Fig.~4
shows that both HFLT and the homogenous SC states expand in $H$-$T$ phase space with Hg doping. For
non-magnetic Sn-doping, on the other hand, both SC and HFLT states contract, and both $T_c$ and $T_{\rm
HFLT}$ at 11\,T reduce with doping, as shown in Fig.~3.


\begin{figure}[t]
\includegraphics[width=\linewidth,keepaspectratio]{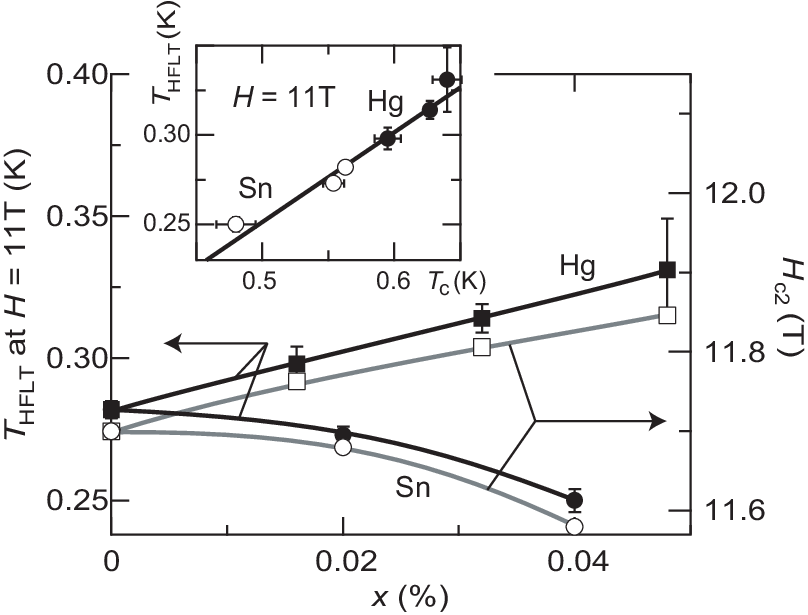}
\caption{Doping dependence of HFLT-phase transition temperature $T_{\rm HFLT}$ at 11\,T for CeCo(In$_{\rm 1-x}$Hg$_{\rm
x}$)$_5$ and CeCo(In$_{\rm 1-x}$Sn$_{\rm x}$)$_5$ (solid symbols, left axis), together with
superconducting upper critical field, $H_{\rm c2}$, of CeCo(In$_{\rm 1-x}$Hg$_{\rm x}$)$_5$ and
CeCo(In$_{\rm 1-x}$Sn$_{\rm x}$)$_5$ (open symbols, right axis). ($\blacksquare$, $\Box$ - CeCo(In$_{\rm
1-x}$Hg$_{\rm x}$)$_5$; $\medbullet$,\,$\medcirc$ - CeCo(In$_{\rm 1-x}$Sn$_{\rm x}$)$_5$.  Solid lines
are guides to eye. Inset: $T_{\rm HFLT}$ vs $T_{\rm c}$ at 11\,T with doping concentration $x$ as an
implicit parameter. Data for both M=Hg and Sn are plotted. Solid straight line goes through the origin.} \label{}
\end{figure}

The correlation between $H_{\rm c2}$ and $T_{\rm HFLT}$ is clearly seen in Fig.~5, where these
quantities are plotted as functions of both Sn- and Hg-doping. Both $H_{\rm c2}$ and $T_{\rm HFLT}$
increase with Hg-doping, while they decrease with Sn-doping. We also plot  $T_{\rm HFLT}$ against
$T_{\rm c}$ at 11\,T in the inset of Fig.~5, with the doping concentration $x$ as an implicit parameter.
Remarkably, a line, which goes through the origin, gives a good fit (solid line).

The observed strong correlation between $T_{\rm HFLT}(H)$ and $T_{\rm c}(H)$ should not be surprising,
since the magnetic HFLT phase only exists within the superconducting state, and therefore its properties
should be strongly linked to the superconducting properties of CeCoIn$_5$. The broadening and
suppression of the anomaly, however, is much more significant with respect to the possible origin of the
HFLT phase. Recent theoretical work~\cite{ikeda-prb-10} on the effect of impurities on both FFLO state
and the magnetic-origin scenario~\cite{Kenzelman-Science09} showed that the FFLO state is much more
fragile to impurities, a result of the softness of the nodal planes due to a strong Pauli
depairing effect in CeCoIn$_5$. Contrary, for a low temperature state with magnetism as a dominant
driving force, one would expect suppression of $T_c$ to zero, without observed strong broadening and
suppression on the specific heat anomaly itself~\cite{ikeda-prb-10}.

In conclusion, we have studied the effect of magnetic (Cd and Hg) and non-magnetic (Sn) doping effects
on the high field - low temperature superconducting HFLT phase of CeCoIn$_5$ by means of specific heat
measurements. $H_{\rm c2}$ increases anomalously with hole doping, while it decreases with electron
doping. The change of $H_{\rm c2}$ is much greater than can be expected to be due to an effective
negative pressure effect, and is dominantly an effect of carrier doping on the Fermi surface. The HFLT
phase is extremely sensitive to Sn-doping with a critical concentration of 0.08\,\%, at most, and is
similar to 0.05\,\% for Hg-doping. We note here that, despite the existence of the long range magnetic
order in the HFLT state, it is destroyed by both the AFM- and non-AFM-inducing impurities. The scaling
of $T_{\rm HFLT}$ against $T_{\rm c}$ at 11\,T indicates a close correlation between the HFLT and the
homogenous SC states. More significantly, the broadening of the HFLT specific heat anomaly and its
suppression with minute amount of impurities supports the FFLO origin of the HFLT state.

We are grateful to I. Vekhter, L. Boulaevskii, M. Graf, J. Sauls, and A. Balatsky for stimulating
discussions. Work at Los Alamos National Laboratory was performed under the auspices of the U.S.
Department of Energy.  A. D. Bianchi received support from NSERC (Canada), FQRNT (Qu\`{e}bec), and the
Canada Research Chair Foundation. Z. Fisk acknowledges support NSF grant NSF-DMR-0854781.

\end{document}